# Inferring synthetic lethal interactions from mutual exclusivity of genetic events in cancer

Sriganesh Srihari[1], Jitin Singla[2], Limsoon Wong[3*] and Mark A. Ragan[1*]

[1] Institute for Molecular Bioscience, The University of Queensland, St. Lucia, Queensland 4072, Australia

[2] Department of Computer Science and Engineering, Indian Institute of Technology Roorkee, Uttarakhand 247667, India

[3] Department of Computer Science, National University of Singapore, Singapore 117417, Singapore

*Correspondence should be addressed to Mark A. Ragan: m.ragan@uq.edu.au | +61 07 3346 2616 or Limsoon Wong: wongls@comp.nus.edu.sg | +65 6516 2902.

## Abstract

**Background:** *Synthetic lethality* (SL) refers to the genetic interaction between two or more genes where only their co-alteration (*e.g.* by mutations, amplifications or deletions) results in cell death. In recent years, SL has emerged as an attractive therapeutic strategy against cancer: by targeting the SL partners of altered genes in cancer cells, these cells can be selectively killed while sparing the normal cells. Consequently, a number of studies have attempted prediction of SL interactions in human, a majority by extrapolating SL interactions inferred through large-scale screens in model organisms. However, these predicted SL interactions either do not hold in human cells or do not include genes that are (frequently) altered in human cancers, and are therefore not attractive in the context of cancer therapy.

**Results:** Here, we develop a computational approach to infer SL interactions directly from frequently altered genes in human cancers. It is based on the observation that pairs of genes that are altered in a (significantly) *mutually exclusive* manner in cancers are likely to constitute lethal combinations. Using genomic copy-number and gene-expression data from four cancers, breast, prostate, ovarian and uterine (total 3980 samples) from The Cancer Genome Atlas, we identify 718 genes that are frequently amplified or upregulated, and are likely to be synthetic lethal with six key DNA-damage response (DDR) genes in these cancers. By comparing with published data on gene essentiality (~16000 genes) from ten DDR-deficient cancer cell lines, we show that our identified genes are enriched among the top quartile of essential genes in these cell lines, implying that our inferred genes are highly likely to be (synthetic) lethal upon knockdown in these cell lines. Among the inferred genes include tousled-like kinase 2 (*TLK2*) and the deubiquitinating enzyme ubiquitin-specific-processing protease 7 (*USP7*) whose overexpression correlate with poor survival in cancers.

**Conclusion:** Mutual exclusivity between frequently occurring genetic events identifies synthetic lethal combinations in cancers. These identified genes are essential in cell lines, and are potential candidates for targeted cancer therapy.

# 1      Background

Cells have evolved to ensure their *viability*. Although typically associated with survival (*i.e.* the ability to maintain homeostasis but not necessarily cell division), cell viability can be defined more broadly to encompass the ability to grow and proliferate. Processes within the cell ensure that it is sufficiently protected against deleterious *genetic events* – *e.g.* mutations, amplifications and deletions – that impact cell viability, but when these events are unavoidable the cell commits to *apoptosis* or programmed cell death.

Genetic events can modify this control on cell viability, resulting in viability being enhanced (*e.g.* in cancer) or compromised (*e.g.* during cell senescence and death). This is effected by the (over-)activation or inactivation of genes responsible for cell viability through *gain-of-function* or *loss-of-function* genetic events, respectively.

When two or more of these genetic events occur simultaneously, these can considerably impact the viability of cells. *Synthetic lethality* (SL), first defined by Bridges in 1922 [1], refers to one such combination between two genetic events (typically affecting two different genes) in which their co-occurrence results in severe loss of viability or death of the cell, although the cell remains viable when only one of the events occurs [2,3].

SL has gained considerable attention over the last few years due to its value in understanding the essentiality of genes or their combinations [4,5], and more recently due to its promise as a therapeutic strategy for selective targeting of cancer cells [6,7]. Cancer cells are genetically different from normal cells and harbour genetic events in specific genes that enhance their viability. Therefore, by identifying and targeting (*i.e.* inducing a genetic event in) the synthetic-lethal partner of these genes, selective killing of cancer cells can be achieved while sparing the normal cells. SL-based therapies exploit these genetic differences in a way that is often not possible with conventional chemotherapy, which is often cytotoxic to normal as well as cancer cells [8].

A pioneering breakthrough in SL-based cancer therapy showed that inhibition of poly(ADP-ribose) polymerase (PARP) in cancer cells that harbour loss-of-function events in the breast-cancer susceptibility genes *BRCA1* and *BRCA2* is dramatically lethal to these cells [9,10] (reviewed in [11]). Germline losses in *BRCA1*/*BRCA2* are highly penetrant, conferring 60-80% risk of breast and 30-40% risk of ovarian cancers. These losses account for about 10-25% of hereditary breast and ovarian cancers [11,12].

Following the promise of BRCA-PARP, several studies have explored (computational) identification of SL interactions that could be efficacious in treating cancer. This began with seminal [13-15] and follow-up works [16-18] that studied "cross-talk" between pathways in model organisms including yeast, worm and fruit fly to characterise genetic interactions.

From these studies emerged a *between-pathway model* [13,14] according to which loss of function in only one pathway does not greatly affect cell viability, but the further inactivation of a second parallel or compensatory pathway results in cell death. This model characterised synthetic lethal interactions as genetic interactions between these compensatory pathways.

More-recent studies [19-21] have attempted extrapolation of SL interactions from model organisms (*e.g.* yeast http://drygin.ccbr.utoronto.ca/ [22] using protein-sequence homology to infer interactions in human cells – *e.g.* BRCA2-RAD52 [23], SMARCB1-PSMA4, ASPSCR1-PSMC2 [19] and between FEN1 and SMC3, RNF20, BLM, MRE11A, STAG3, CDC4 and CHTF18 [20,21]. Classification-based approaches [24-27] that employ a support vector machine trained with features from model organisms have also been used to predict new SL interactions in human, with the expectation that SL interactions follow similar organisational principles in human and model organisms. Recently Zhang *et al.* [28] proposed that single- and double-knockdown of proteins within known pathways could be computationally simulated to estimate interactions that are lethal; AKT with BID, CASP9 and WEE1 were among the top SL interactions identified in human. Others [29-33] have employed combined experimental and computational approaches by performing knockdown of combinatorial pairs of genes using large-scale siRNA-screens across cell lines and *in vivo* models (*e.g.* http://www.genomernai.org/ [30]). These approaches have been more successful than the solely computational ones, resulting in identification of actionable SL-based targets, including GATA2 and CDC6 as SL partners of KRAS [29]. However, these approaches are considerably more expensive, and many of the essential genes so identified turn out to be either restricted to only these cell-line models or are infrequently overexpressed in cancers.

Despite these attempts, interactions extrapolated from lower-order model organisms fail to hold up in human cells and are less-appealing in the context of cancer therapy. This is because the model systems, despite sharing some homologous proteins with human, have considerably different and simpler cellular and functional organisation [34,35]. While core cellular processes including cell-cycle and DNA-damage repair are broadly conserved, human cells express novel proteins, isoforms and/or paralogs with partially overlapping functions that buffer the loss of one another [34-37], with the consequence that lethality inducible by targeting only one of them is not conserved. For example, human cells have three AKTs – AKT1, AKT2 and AKT3 – with partially overlapping functions [38] whereas yeast has only one AKT. Similarly, SL interactions predicted from 'static' pathway maps do not reflect the actual scenario in cancer cells; these cells undergo significant pathway rewiring to enhance their viability [39,40]. Finally, these SL interactions do not include cancer genes or genes that are frequently altered in cancers – in particular, from the examples above, *ASPSCR1*, *BLM*, *SMARCB1* and *MRE11A* put together are altered (homozygous deletion) in < 10% of most cancers as per The Cancer Genome Atlas (TCGA) Cbioportal cohort [41,42] – and therefore, the proportion of cancers benefiting from targeting their SL partners is very small.

Here, we develop a computational approach taking into account the above factors by directly inferring SL interactions from frequently altered genes in cancers. We show that specific combinations of genes that display *mutual exclusivity* for genetic events are likely to constitute lethal combinations, and therefore by targeting these genes in *conjunction* could kill cancer cells. To demonstrate this, we consider six key DNA-damage response (DDR) genes that are frequently altered across four cancers – breast, prostate, ovarian and uterine – and using genomic copy-number and gene-expression data from TCGA [41,42], we identify genes that are altered in a (significantly) mutually exclusive manner with these six DDR genes. By comparing with data from genome-wide (~16000 genes) essentiality screens across ten DDR-deficient cancer cell lines [31,32], we show that our identified genes are enriched among the top quartile of essential genes in these cell lines, implying that our inferred genes are likely to be (synthetic) lethal upon knockdown in these cell lines.

## 2  Methods

Suppose that in a given large set of viable (cancer) cells, a pair of genes exhibits *mutual exclusivity* with respect to a genetic event – *i.e.* each gene individually is affected by the genetic event in most large proportions of cells but both genes are simultaneously affected in few or none. We hypothesize that the observed viability of these cells is dependent on, or a consequence of, the mutual exclusivity between the two genes: cells are not viable if the genetic event were to affect both genes simultaneously, and therefore we observe that few if any viable cells in our population carry such an event in both genes (in other words, the mutually exclusive combinations constitute the (clonally) selected combinations amenable to cell survival). Consequently, we infer that the two genes are synthetic lethal with each other.

### 2.1  *Mutual exclusivity between genetic events and inferring SL combinations*

Suppose that we are given an arbitrarily large set of viable (cancer) cells $S$. Let $A$ and $B$ be a pair of genes affected by a genetic event $E$ in these cells $S$. Let $S_A$ (respectively, $S_B$) be the subset of $S$ in which $A$ (respectively, $B$) is affected by $E$, and let $S_{AB} = S_A \cap S_B$. The mutual exclusivity between $A$ and $B$ with respect to $E$ can be defined as both $|S_A|/|S_{AB}|$ and $|S_B|/|S_{AB}|$ approaching infinity as $|S|$ approaches infinity. Given this mutual exclusivity we infer that $A$ and $B$ are synthetic lethal with each other.

**Claim:** $|S_A|/|S_{AB}|$ and $|S_B|/|S_{AB}|$ both approach infinity as $|S|$ approaches infinity if, and only if, the co-occurrence of $E$ in $A$ and $B$ affects cell viability, and therefore $A$ and $B$ are synthetic lethal with each other.

**Basis for the claim:** Suppose the co-occurrence of $E$ in $A$ and $B$ is lethal to the cells but not in either A or B alone. Then, $S_{AB} = \phi$ or a very small proportion of $S$. Therefore, $|S_A|/|S_{AB}|$ and $|S_B|/|S_{AB}|$ approach infinity when $S$ is large.

Conversely, $|S_A|/|S_{AB}|$ and $|S_B|/|S_{AB}|$ both approaching infinity implies that as event $E$ in $A$ occurs more often and $E$ in $B$ occurs more often, the co-occurrence of $E$ in $A$ and $B$ occurs less often. But, if $A$ and $B$ are independent, we do not expect to see this pattern. So, the occurrences of E in A and B are avoiding each other in viable cells, and thus their co-occurrence affects cell viability.

*2.2 Computing significant mutually exclusive gene combinations*

Let *X* be the random variable that counts the number of cells that show co-occurrence of a genetic event for the pair of genes (*A*, *B*). We estimate the statistical significance for the mutual exclusivity between *A* and *B* based on the probability of observing at most $|S_{AB}|$ cells (out of $|S_B|$ cells) with co-occurrence of the event (with $|S_A|$ cells). We estimate this probability $P[X \leq |S_{AB}|]$ as

$$P[X \leq |S_{AB}|] = 1 - P[X > |S_{AB}|], \quad \text{(Equation 1)}$$

where $P[X > |S_{AB}|]$ is computed using the hypergeometric probability mass function for $X = k > |S_{AB}|$:

$$P[X > |S_{AB}|] = \sum_{k=|S_{AB}|+1}^{|S_B|} \frac{\binom{|S_A|}{k}\binom{|S|-|S_A|}{|S_B|-k}}{\binom{|S|}{|S_B|}}$$

This "1 – hypergeometric test" *p*-value (Equation 1) is used to infer SL pairs (at $p < 0.05$), and the inferred pairs are ranked in order of their *p*-values.

## 3    Results

*3.1    Datasets*

We gathered genomic copy-number and gene-expression datasets from four sporadic cancers, breast [43], prostate [44], ovarian [45] and uterine [46], from TCGA *via* Cbioportal (http://www.cbioportal.org/index.do)   [41,42]   and   TCGA   Firehose (http://gdac.broadinstitute.org/), composing a total of 3980 samples (Table 1). We consider four distinct genetic events: two kinds of *genomic* events (*viz.* gene copy-number amplifications and deletions), and two kinds of *expression level* events (*viz.* gene up- and downregulation). We expect that the changes in expression levels should encompass the effects of other kinds of events not directly considered here – *e.g.* mutations, chromatin changes and methylation.

These copy-number and expression events are inferred from GISTIC-normalized [47] values available *via* Cbioportal [41,42] and TCGA Firehose. The copy-number value for each gene reflects the deviation in its number of copies from normal and is normalized to a range of [-2,

2] where negative values represent deletions and positive values represent amplifications. We consider only high-level amplifications and deletions (typically homozygous deletions) having copy-number values +2 and -2, respectively. Likewise, the expression for each gene is *z*-score normalized, and here we consider genes that are highly upregulated or downregulated given by *z*-scores at least two standard deviations on either side of the mean (as per [41,42]). For more details on how these GISTIC-normalized values are computed, refer to [47].

To validate our predictions (genes *B*) we employed genome-wide (~16000 genes) essentiality data from siRNA-mediated knockdown screens across ten cancer cell lines that harbour a deficiency (mutation, deletion or downregulation) in at least one of the genes *A* (Table 2) [31,32]. These essentiality data are in the form of GARP (Gene Activity Rank Profile) scores for each gene and are approximately in the range [+5, -10] with a lower value in a cell line indicating higher essentiality for the gene in that cell line.

*3.2   Identifying mutually exclusive combinations involving frequently altered genes in cancers: a case study using six DNA-damage response genes*

Here we consider mutual exclusivity with deletion and downregulation events affecting the following six genes (as genes *A*): *ATM*, *BRCA1*, *BRCA2*, *CDH1*, *PTEN* and *TP53*. These are tumour-suppressor genes that are central to or regulate DNA-damage response (DDR) functions, that is, genes that play important roles in maintaining the genomic integrity of the cell and control cell proliferation [11]. These genes are deleted or downregulated across all the four cancers considered here, and their loss is a significant driver event in these cancers (TCGA, 2011; TCGA, 2012; TCGA, 2013; TCGA, 2014). To predict SL interactions, we identify genes *B* that are either amplified/upregulated or deleted/downregulated in a mutually exclusive manner to these genes *A*. Specifically, we identify two kinds of mutually exclusive combinations (SL interactions): (i) deletion/downregulation of gene *A* with amplification/upregulation of gene *B*; and (ii) deletion/downregulation of gene *A* with deletion/downregulation of gene *B*.

*3.2.1 Gene A deletion or downregulation with gene B amplification or upregulation*

We identified a total of 842 SL interactions involving 718 genes *B* at $p<0.01$. Figure 1a shows the distribution of these interactions with respect to the different genes *A*. *BRCA2* dominates the number of SL interactions followed by *CDH1*, *PTEN* and *TP53*, whereas *ATM* and *BRCA1* participate in very few SL interactions. While these proportions are to an extent influenced by the actual fraction of cases in which these genes *A* are deleted/downregulated – *PTEN* (34%), *CDH1* (12.9%), *BRCA2* (9.8%), and *TP53* (9.7%) are in much higher numbers than *ATM* (5.7%) and *BRCA1* (4.8%) across all cancers – this still indicates that overall (*i.e.*

across the four cancers) these six genes are involved to different extents in their synthetic lethality with amplification/upregulation of genes *B*. Moreover, there were fewer (<5%) overlaps between genes *B* partnered with different genes *A*, indicating considerable diversity in the SL landscape (Additional file 1).

However, different genes *A* dominate the SL interactions within the individual cancers – *e.g.* *CDH1* (99.5%) for breast, *PTEN* (78.4%) and *BRCA2* (16.8%) for prostate and *BRCA1* (17.9%) and *TP53* (16.2%) for ovarian cancers at $p < 0.01$, as shown in Figure 1b. These results indicate that SL interactions could be highly context-dependent (here, the type of cancer) with deletion/downregulation in different genes *A* dominating SL interactions within different cancers. Note that these analyses are based on uncorrected *p*-values (Equation 1) and are meant only to give a sense of the distribution of SL interactions; for the validation of genes *B* (below) we use the relative rankings of their *p*-values.

*3.2.2 Gene A deletion or downregulation with gene B deletion or downregulation*

The number of SL interactions involving both genes *A* and *B* deleted/downregulated were considerably fewer than in the previous case – a total of 143 interactions involving 117 genes *B* identified across all the four cancers at $p < 0.05$. As above, *BRCA2*, *PTEN* and *CDH1* dominate these SL interactions (Figure 2a) when all four cancers are taken together, whereas different genes dominate within the individual cancers – *e.g. CDH1* (91%) and *PTEN* (9%) for breast, *PTEN* (88%) and *BRCA2* (9%) for prostate, and *CDH1* (58%) and *BRCA1* (38%) for ovarian cancers.

*3.3    Computational validation using data from cell-line essentiality screens*

We expect that targeting genes *B* in conjunction with genes *A* could induce lethality. To validate this, we analysed the GARP essentiality scores of genes *B* in cell lines deficient with genes *A*. We chose ten cell lines (Table 2) that harbour a deficiency in at least one of the genes *A* [31,32]. The left-hand side plots of Figure 3 compare the ranges of GARP essentialities of our predicted genes *B* with that of the entire set of ~16000 profiled genes in these cell lines. While it is difficult to directly compare the two ranges because of the difference in number of genes in them, for the majority of cell lines the genes *B* at the $25^{th}$ percentile had lower GARP scores than the corresponding genes from the entire profiled set. In particular, our predicted genes *B* were enriched significantly ($\chi^2$ test $p<10^{-5}$) with the top-quartile of essential genes (approximately the top 5000) from these ~16000 genes.

We ranked our predicted genes *B* in increasing order of their mutual-exclusivity significance to generate a *mutual-exclusivity (ME) ranking*. Then, for each gene *B* that was among the top

5000 we assigned a *gene-essentiality* (*GE*) *rank* as '5000 – rank of gene *B* in the essentiality screen' (a reverse ranking). We then plot GE rank *vs* ME rank for all genes *B* for each cell line according to the gene-*A* deficiency it harbours. For example, since the cell line HCC1143 harbours a deficiency in *TP53* (Table 1), we plot GE rank *vs* ME rank for all genes *B* that are predicted as mutually exclusive with *TP53* using the GARP score data for HCC1143. Doing so using the amplified/upregulated genes *B* resulted in the right-hand side plots shown in Figure 3. For all cell lines harbouring deficiencies in *ATM*, *BRCA1*, *BRCA2*, *PTEN* and *TP53* we see a downward trend, thereby indicating a strong agreement between the rankings based on mutual exclusivity and the GARP essentialities of our predicted genes *B* in gene *A*-deficient cell lines (no data are available for cell lines with *CDH1* deficiency). This analysis indicates that the genes *B* that are mutually exclusive with the loss of *A* in tumours are also essential in gene *A*-deficient cell lines, and therefore supports our hypothesis that the observed mutual exclusivity is very likely a mechanism to avoid cell lethality (thereby enhancing the essentiality of *B*). As these genes *B* are also (frequently) amplified/upregulated in the cancers, these could be attractive as targets in cancer therapy.

Figure 4 shows similar plots using the deleted/downregulated genes *B*; however, since these genes are far fewer the plots show data for fewer gene ranks, the most being for *PTEN*.

To understand whether the lethality observed for genes *B* is specific to *A*-deficient cell lines, we analysed the differential essentiality of *B* in the cell lines relative to the MCF7 cell line (due to lack of suitable data on normal cell lines, we chose MCF7 which is a typical luminal line with no known DDR defect, as our control for the comparison). We observed significant difference between the mean essentialities for *B* between the DDR-deficient and MCF7 cell lines (Figure 5a). Similar results were observed using data [32] from two HCT116-derived isogenic cell lines, one *PTEN*-/- and the other with wild-type *PTEN* (Figure 5b). This analysis indicated that the essentiality of *B* was highly specific to cell lines harbouring gene *A* deficiency, and hence *B* is synthetic lethal in the context of deficiencies in DDR genes.

### 3.4 Case studies of identified SL partners B

We expect that our predicted genes *B* that are amplified/upregulated to confer poor survival, and to validate this we plot the Kaplan-Meir curves for these genes using survival data from cancer patients (KMPlotter: http://www.kmplot.com/) [48]. Figure 6 shows examples for breast cancer (plots for ovarian cancer in Additional file 1).

Several interesting genes are identified here – *e.g. TLK2*, which encodes the serine/threonine tousled-like kinase, is closely associated with the repair of DNA double-strand breaks (DSBs) and in the regulation of chromatin assembly during the S-phase (http://www.genecards.org/ ) [49]. *TLK2* is amplified/upregulated in 26% of sporadic breast cancer cases, and confers significantly poor survival ($p$ = 0.00072). In particular, GOBO-based analysis (http://co.bmc.lu.se/gobo/gobo.pl) [50] indicates that *TLK2* is upregulated in 37% luminal

(estrogen receptor (ER)-positive) breast tumours with grade 3-stratified multivariate analysis showing a hazard ratio of 2.25 ($p=10^{-5}$) and poor survival ($p<0.01$) in these tumours (Additional file 1).

Several interesting genes are identified here – *e.g. TLK2*, which encodes the serine/threonine tousled-like kinase, is closely associated with the repair of DNA double-strand breaks (DSBs) and in the regulation of chromatin assembly during S-phase [49]. *TLK2* is amplified/upregulated in 26% of sporadic breast cancer cases, and its amplification/upregulation is mutually exclusive to *BRCA2* deletion/downregulation. *TLK2* overexpression correlates with significantly poor survival ($p = 0.00072$) in these patients. In particular, GOBO-based analysis [50] indicates that *TLK2* is overexpressed in 37% luminal (estrogen receptor (ER)-positive) breast tumours, and grade 3-stratified multivariate analysis indicates a hazard ratio of 2.25 ($p=10^{-5}$) and poor survival ($p<0.01$) in patients with these tumours (Additional file 1). Interestingly, *TLK2* overexpression can co-occur with *PTEN* loss or when *PIK3CA*, a key driver of ER-positive/luminal tumours, is not overexpressed. This also agrees with the high expression of *TLK2* in luminal cell lines MCF7, MDA-MB-361 and SUM52PE which do not show high expression for *PIK3CA* (Additional file 1). Therefore, it is possible that *TLK2* acts as a *context-dependent driver* of ER-positive/luminal tumours in the absence of *PIK3CA* expression.

The ubiquitin specific peptidase USP7, which is a deubiquitinating enzyme, is amplified/overexpressed in ~40% of breast tumours in TCGA. USP7 is known to deubiquitinate target proteins including TP53 and PTEN. Overexpression of USP7 correlates with poor survival specifically in *TP53*-mutant patients (Additional file 1).

*EXOSC4* which encodes the EXOSC4 subunit of the RNA exosome complex that is important for RNA processing and degradation, is upregulated in 24% breast tumours, and confers poor survival ($p=0.012$).

## 4 Discussion

Our results demonstrate that there exist pairs (*A*, *B*) of genes that are altered in a mutually exclusive manner across tumours. We hypothesize that this observed mutual exclusivity could be a mechanism to avoid cell death, and consequently these pairs (*A*, *B*) constitute synthetic lethal combinations. To test our hypothesis, we use the essentialities (GARP scores) measured for genes *B* across cell-lines deficient in genes *A* (here, *A* includes six key DDR genes). We demonstrate that when our predicted genes *B* are ranked in order of their mutual exclusivity with *A* (as *p*-values), their ranks are consistent with that of their GARP scores in these *A*-deficient cell lines: the top-ranked genes *B* are also highly essential (lethal) to these cell lines. This is strongly suggestive that mutual exclusivity is an important mechanism to avoid lethality.

Our approach is novel because we infer SL interactions directly from tumour data and validate these using essentiality data from tumour cell lines, and is different from earlier approaches [24-28]. Therefore, we effectively bypass several of the limitations of these approaches *viz.* inference of infrequently altered genes as SL partners, inference of genes solely from cell line models that may not hold in tumours, and inference of SL interactions from model organisms that do not hold in human [35].

Beginning from the between-pathway model [13,14], synthetic lethality (SL) has often been associated with compensatory or parallel pathways, such that the loss of function of one of the pathways does not significantly affect cell viability whereas the loss of both pathways results in cell death (Figure 7a). Although this classical view gives an elegant explanation for SL, it only presents a partial one, mainly in terms of loss-of-function (inactivation) events. However, in general SL could also involve gain-of-functions (activation) events. Moreover, in the context of cancer, this model caters mainly to tumour-suppressor genes. For example, the loss of function in two parallel DNA-damage repair (tumour suppressor) pathways can lead to a considerable accumulation of DNA damage, resulting in genomic catastrophe and triggering apoptosis in cancer cells, as in the case of *BRCA-PARP* [11]. However, tumour-suppressor genes in general can be difficult to target because of the (unknown) side-effects these could have on normal cells [51, 52], and because these are infrequently (over-)expressed in cancers to enable their targeting. Interestingly, many of the SL interactions that are extrapolated from lower-order organisms turn out to be tumour-suppressor genes (*e.g.* *SMARCB1*, see Introduction) and these are rarely altered in human cancers. We suspect that many of these highly conserved genes are much less susceptible to alterations than are newer inventions in humans, and consequently do not form attractive therapeutic targets in human cancers.

Here, we extend these pathway models [13,14] to include gain-of-function (activation) events (Figure 7). In addition to the parallel-pathway model (Figure 7a) we propose a *negative feedback-loop model* (Figure 7b) wherein the forward path involves a gain-of-function event (often in an oncogene) whereas the negative-feedback loop involves a loss-of-function event (often in a tumour-suppressor gene). Cell viability is enhanced by activation events in the forward path or reciprocally by inactivation events in the negative-feedback loop. We hypothesize that, in the event of loss of a DDR gene in the negative-feedback loop, the simultaneous activation of an oncogene in the forward path could be detrimental to the cell's survival by generating genomic instability. Consequently, to maintain an optimal condition for survival, cancer cells harbour only one of the two events resulting in mutual exclusivity between these events. We suspect the *PIK3CA-PTEN* combination is one such case: the *PI3K* pathway either harbours frequent activation events in the oncogenic *PIK3CA* kinase (96/156 breast tumours) resulting in accelerated cell growth and proliferation or reciprocally frequent inactivation events in the tumour suppressor *PTEN* (67/156) resulting in loss of negative-feedback to control cell proliferation; however we rarely see breast tumours harbouring both these events (8/156; *p-value*≈0) possibly due to their detrimental effect on cancer cell survivability.

Likewise, the *KRAS-NF1* combination also fits into this pathway model. Simultaneous activation of the *KRAS* oncogene together with inactivation of *NF1* tumour suppressor could be lethal to cell survival, and hence these two events rarely co-occur (*p-value* = 0.001).

Another example of SL is between *BRCA1* and *CCNE1*, which although are not components of the same physical pathway, are functionally related due to their roles in the cell cycle and therefore broadly fit into our proposed model. *BRCA1*, being a tumour suppressor and a regulator of DNA-damage repair, has a reciprocal role to *CCNE1* whose overactivation accelerates cell divisions and confers replication stress and genomic instability. *CCNE1* amplification/overexpression is mutually exclusive to *BRCA1* deletion/underexpression in ovarian cancers (*p-value* = 0.073). Consequently, loss of *BRCA1* is synthetic lethal to cells harbouring *CCNE1* amplifications, and this has recently been validated using inhibition of BRCA1-mediated DNA repair in ovarian cancer cell lines [53].

While the induction of cell lethality for certain combinations of genetic events seems a compelling reason for the observed lack of tumour samples containing these combinations, an alternative explanation could be that cells use mutual exclusivity as a means to achieve multiplicity in phenotypes. For example, it is possible that *PIK3CA* activation and *PTEN* inactivation are two (disjoint) paths to achieve two distinct phenotypes. One observation in support of this is that *PIK3CA*-activated breast tumours tend to be luminal, whereas *PTEN*-inactivated breast tumours tend to be basal-like [43]; however, harbouring simultaneous events in both genes is not additively advantageous to cells. A similar explanation also underlies the mutual exclusivity for *KRAS* and *EGFR* mutations seen in lung cancer [54].

Another example is *CDH1-PTK2*. Here, the tumour suppressor *CDH1* is responsible for maintaining cell adhesion and regulating cell migration. The focal adhesion kinase *PTK2* is responsible for disassembly of cell adhesions and promoting cell proliferation and migration. *CDH1* inactivation is mutually exclusive to *PTK2* activation in breast cancer (*p-value* = 0.001). During tumour development and in particular during metastasis, the inactivation of *CDH1* or alternatively the activation of *PTK2* could be two disjoint paths to achieve cell migration to distant sites.

*4.1 Genes B as targets in cancer*

Consistent with these pathway models, we expect that targeting *B* in the context of deletion/downregulation of *A* could result in cancer-cell death by either disrupting both survival pathways (Figure 7a) or by shutting off (forward) signals for cell survival (Figure 7b). In the latter case, targeting gene *B* irrespective of the (deletion/downregulation) status of *A* could result in cancer cell death, a scenario referred to as "oncogene addiction" [55,56]. Our proposed model (Figure 7b) subsumes this scenario, and hence presents a more-general strategy for targeting oncogenes under the synthetic lethality paradigm.

# 5    Conclusion

In recent years, SL has emerged as an attractive therapeutic strategy against cancer: by targeting the SL partners of altered genes, cancer cells can be selectively killed while normal cells are spared. Here we introduce a computational approach to infer SL interactions based on the frequency at which genes are altered in human cancers. It is based on the observation that pairs of genes that are altered in a (significantly) mutually exclusive manner in cancers are likely to constitute lethal combinations. Using omics datasets across breast, prostate, ovarian and uterine cancer, we identify 718 genes that are upregulated or amplified in cancers, and are likely to be synthetic lethal with six key DDR genes. Computational validation of our predicted genes using essentiality data from cell-line screens shows that these genes are among the top essential genes and therefore likely to be lethal upon knockdown in these cell lines. We intend to validate some of these genes using single- and double knockdown in cell line and *in vivo* cancer models.

**List of abbreviations**

DDR: DNA-damage response
GARP: Gene Activity Rank Profile
ME: Mutual exclusivity
SL: Synthetic lethality
TCGA: The Cancer Genome Atlas

**Competing interests**

The authors declare that they have no competing interests.

**Authors' contributions**

SS and LW conceived the idea and formalised it. SS and JS performed the analyses. SS drafted the manuscript and the response to reviewers' comments. JS, LW and MAR contributed to writing and revising the manuscript. All authors have read and approved the final manuscript and response letter.

**Acknowledgements**

We thank Professor Kum Kum Khanna and Dr Peter Simpson for valuable discussions. This work is supported by Australian National Health & Medical Research Council project grants 1028742 and 1080985 to Dr Peter T. Simpson, Dr Nicola Waddell and MAR. A preliminary version of this work will be presented as a poster discussion at the San Antonio Breast Cancer

Symposium 2015, and SS thanks the Ian Potter Foundation (Australia) for travel support (grant# 20160303).

# RESPONSE TO REVIEWERS

## Reviewer 1 (Dr Michael Galperin)

*I support publication of this manuscript in its current form.*

Thank you for reviewing our manuscript and for your supportive comment.

## Reviewer 2 (Dr Sebastian Maurer-Stroh)

*The manuscript 'Inferring synthetic lethal interactions from mutual exclusivity of genetic events in cancer' is an interesting study outlining a new computational statistical approach to identify synthetic lethal (SL) pairs for potential targeting in cancer treatment from large-scale gene copy number and expression data with example validation against essentiality data from siRNA screens. The introduction gives an adequate overview of prior related work and the background context. The new approach hinges on the observation that there seems to be an inverse correlation between gene essentiality and mutual exclusivity of co-occurrence of genomic events with mutated tumor suppressor genes in the studied cell lines (right side panel of Figure 3). I have two main comments:*

*1) While the correlation in Figures 3 and 4 (right panel) is certainly interesting, I am wondering how SPECIFIC the identified SL pairs are in these cases to the actual gene pairs. The Figure only shows data/lines for cell lines that have a defect in the respective DDR gene A for which the mutual exclusivity with genes B is calculated. It would be important to show also the data/lines for all other cell lines used here (maybe in weak gray color so they stay in the background) so one can see if the correlation is specific to the studied SL pair with gene A or possibly a general function of the essentiality of genes B. The latter could be seen if a similar correlation would appear independent of defect status of gene A. Pair specificity would be important for future clinical use since a specific link would mean that a genetic test for defects in gene A would be useful, while unspecific linkage would not require it.*

Thank you for this important comment. Certainly the specificity of our identified SL targets, or in other words, the differential essentiality/lethality of these targets in DDR-deficient cell lines *vis-à-vis* normal cell lines or in cell lines with proficient DDR, is an important concern. Due to lack of suitable data from normal cell lines, here we analyzed the differential essentiality of these genes relative to the MCF7 cell line, which is a typical luminal cell line that does not harbor any known DDR defect. In addition, we compared the essentialities of these genes between *PTEN* wild-type and *PTEN*-/- isogenic HCT116-derived cell lines (obtained from [32]) to understand differential essentiality in the context of *PTEN* loss. In both analyses, we observe significantly higher essentialities for our genes in DDR-deficient cell lines, thereby indicating that knockdown of these genes is more lethal specifically in the context of DDR defects. These analyses have now been added to **Section 3.3** and as a new **Figure 5**.

*2) It would be a stronger argument for selected SL pairs if they can be mechanistically rationalized and discussed to a greater extent in their pathway context (e.g. double hit in same pathway or hit in two different redundant pathways). Maybe add 1 or 2 more explicit examples like the one for PIK3CA-PTEN in the discussion.*

This is again a very useful suggestion. We now discuss three other examples in the manuscript – *BRCA1-CCNE1*, *KRAS-NF1* and *CDH1-PTK2* under **Discussion** – to provide mechanistic explanation to our hypothesis.

**Reviewer 3 (Professor Sanghyuk Lee)**

*The authors propose an intriguing hypothesis that synthetic lethality underlies the mutual exclusivity in cancer. Using the TCGA data sets of four cancers (3980 samples), they identified 718 genes that were mutually exclusive with six key DNA-damage response genes. Their rank correlation with the synthetic lethality measured in ten DDR-deficient cancer cell lines showed a strong linear relationship. Survival plots for several candidate genes showed significant separation, further supporting the hypothesis.*

*The concept of inferring synthetic lethality from the mutual exclusivity is novel even though several studies (Ciriello G et al., Genome Research 2012; Unni AM et al., eLife 2015) asserted those two are related. Systematic mining of the TCGA data sets provides a solid ground for the hypothesis.*

*There are several concerns that should be addressed:*

*1. The box plots are not helpful at all. The authors need a better illustration showing that mutually exclusive genes were enriched in top quartile of essential genes. The rank correlation plot is confusing as well because of the reverse ranking for the gene essentiality. It would be more intuitive to take the positive correlation in the plot.*

The huge difference between the sizes of the two gene lists – our set of 718 genes *vs* ~16000 profiled genes – makes it difficult to have a better illustration for the side-by-side comparison between their GARP values. But, we mention this now in the caption of **Figure 3**. Hence, we report the Chi-square test *p*-value ($p<10^{-5}$) for the enrichment of our genes in the top-quartile of all profiled genes across the different cell lines.

We agree that showing the positive correlation might look more natural, but then if we were to plot using increasing ranks, we won't know at which rank to stop. Therefore, here we put a cut-off and begin at rank 5000 (because this covers the top-quartile of the ~16000 genes into which our predicted genes mostly fall), and then go down until rank 0. This inverse plot helps to convey the idea that as we go down the ME ranks (*i.e.* with higher co-occurrence) the essentiality of gene *B drops* in cell lines.

*2. The authors show the Kaplan-Meier curves for candidate genes. It is not clear how the patients were divided into two groups of high and low. KM plots tend to depend on the choice of patient cohorts dramatically. The same tendency should be replicated using independent patient cohorts. I assume that there exist several well-known cohorts of breast cancer with gene expression data available in public.*

The patients were divided into two groups based on the expression levels of the genes in question, as patients overexpressing the genes (expression levels in the upper tertile) and patients underexpressing the genes (expression levels below the upper tertile). This information is now added to the **Figure 6** caption.

We used KM Plotter [57] to plot the KM curves. KM Plotter already combines multiple cohorts covering more than 1800 patients to generate the curves. For some of the genes, we also used GOBO [50] which also combines several cohorts.

*3. To enhance the reliability of the assumption (i.e. synthetic lethality directly related to mutual exclusivity), it would be of great help to show well-known examples in the introductory part. EGFR-KRAS in lung cancer was highlighted by a recent paper (Unni AM et al., eLife 2015). I expected a*

*similar trend for the BRCA-PARP genes in breast cancer, but was puzzled to find that they do not seem to be mutually exclusive in a simple query to the cBioPortal. Do authors have any explanation? What happens to the MDM2-TP53 gene pair (Ciriello G et al., Genome Research 2012)?*

Synthetic lethality is a broader concept and in general covers any combination of genetic events that induces lethality in cells. Here, we only analyze expression up/downregulation and genomic amplification/deletion events affecting six DDR genes to study SL in the context of DDR defects. Our rationale behind focusing on DDR is that cells already deficient in DDR functions would be highly sensitive to further DNA damage, and therefore any event that triggers further accumulation of DNA damage – the upregulation/amplification of an oncogene that results in fast cell divisions, or the downregulation/deletion of a compensatory DDR gene that further weakens DNA-damage repair functions – could be highly lethal to the cells (genomic catastrophe). By basing on mutual exclusivity, we present a systematic approach to mine for SL cases, and demonstrate our approach on DDR genes; but the SL cases reported in the literature covering non-DDR genes might not be detected from our analysis.

Moreover, lethality depends on the functions and essentialities of the affected genes – *e.g.* events affecting two genes might be avoiding one another (*i.e.* mutually exclusive) as a means to increase the number of possible paths to achieve distinct phenotypes, but their individual or simultaneous occurrence might not necessarily be lethal to the cells (see **Discussion**). Furthermore, several confounding factors – *e.g.* the existence of other events that rescue cells from lethality – could also affect these SL relationships. Some examples including BRCA-PARP and how cells are rescued from lethality *via* loss of 53BP1 are discussed below to further highlight these intricacies.

The BRCA-PARP combination is the first successful example of a SL relationship that has seen clinical applicability in breast and ovarian cancers, to kill cancer cells deficient in BRCA1/2 (cells with 'BRCAness' property). Here, the PARP1 protein is a key component of the alternative non-homologous end-joining (NHEJ) pathway which functions throughout the cell-cycle to repair double-strand breaks (DSBs). PARP1 also plays an important role in restarting of stalled forks during DNA replication (S-phase) by directing DSBs (converted from stalled forks) towards BRCA-mediated homologous recombination (HR) pathway for DSB repair. The HR pathway mainly functions during the G1/S phase to repair DSBs arising during DNA replication.

Consequently: (i) PARP, being part of the NHEJ machinery, functions throughout the cell cycle, and in general is overexpressed in fast-dividing cancer cells that harbour DNA breaks; but (ii) the functions of PARP and BRCA1-mediated repair overlap specifically during the DNA replication (G1/S-phase). In general, any query on cancer datasets (including Cbioportal for TCGA) would show

PARP1 overexpressed in a considerable proportion of tumours. But, the synthetic lethality between BRCA1 and PARP applies to the small window during DNA replication where their functions become compensatory: in the event of loss of BRCA1, further inhibition of PARP results in accumulation of DSBs (because DSBs are not repaired or are repaired erroneously), thus leading to accumulation of lethal levels of DNA damage and consequently cell death. Therefore, PARP overexpression is not necessarily mutually exclusive to BRCA1 loss (*i.e.* PARP is not overexpressed only to compensate for BRCA1 loss), but their double loss is catastrophic and hence synthetic lethal to cells.

Intriguingly, reports have also shown that an additional loss of *53BP1* can rescue cells BRCA-deficient cells from lethality despite PARP inhibition [58]. This is an excellent example of where additional genetic events act as confounding factors to SL relationships, and is also suspected to be a mechanism to acquire drug resistance in cancers.

Another interesting example is the synthetic lethality upon inhibition of checkpoint kinase *CHEK1* and cyclin-dependent kinase *CDK1* in cells overexpressing the *MYC* oncogene [59-61]. Cells overexpressing *MYC* display accelerated cell divisions, and the resulting replication stress leads to accumulation of significant levels of DNA damage. Cell-cycle checkpoints arrest these cells with DNA damage and give them time to repair their damaged DNA. Inhibition of *CHEK1*/*CDK* results in the abrogation of cell-cycle checkpoints, and consequently the cells are forced to pass through checkpoints unstopped, leading to a genomic catastrophe and cell death. Here again, if we query cancer datasets, *CHEK1*/*CDK* being cell-cycle regulators, are seen overexpressed in most aggressive tumours (*e.g.* triple-negative breast cancers), and but are not necessarily mutually exclusive to *MYC* overexpression. However, *CHEK1*/*CDK* inhibition proves catastrophic to the fast-growing cells, and in that way *CHEK1*/*CDK* inhibition is lethal to cells overexpressing *MYC*.

MDM2 is a negative regulator of TP53, and its overexpression inactivates TP53 functions. Consequently, even if some cancer cells have wild-type TP53, the overexpressing MDM2 prevents activation of TP53-mediated apoptotic program. Hence, inhibition of MDM2 releases the brakes on TP53, thus enabling cells to be arrested and driven down the path of programmed cell death. Therefore, here again, TP53-MDM2 does not necessarily show up as mutually exclusive in cancer datasets, but MDM2 inhibition induces lethality in cells by reactivating the TP53-mediated apoptotic program.

*EGFR-KRAS* is a slightly different story. Here, cancer cells either harbour either *EGFR* mutations or alternatively *KRAS* mutations to confer tumorigenesis, but simultaneous events in both *EGFR* and *KRAS* are not additively advantageous to the cells, as also explained in Unni *et al.* [54]. Mutations in *EGFR* and *KRAS* might be two different ways of conferring cancer phenotypes, but simultaneous mutations in both the genes might be non-optimal for cancer cell survival, and in this way, mutations in *EGFR* and *KRAS* are synthetic lethal. This is now covered under **Discussion**.

In summary, while synthetic lethality provides a logical explanation for the lack of certain co-occurring (*i.e.* mutually exclusive) combinations, in general synthetic lethality is a broader concept. Moreover, most of the SL-based targets observed in the literature are pharmacologically induced (*e.g. via CHEK1* inhibition), but alterations to these targets are not naturally occurring in tumours (*CHEK1* is not frequently lost in tumours), and this explains why some of these genes highlighted in the literature do not turn up in our list.

*4. According to the authors' pathway models, I expect that the amplification/upregulation candidates (718 genes identified) would be enriched with oncogenes. The list of oncogenes can be obtained from various resources including the Cancer Gene Census, Vogelstein's paper on 20/20 rules, and so on. This simple test would enhance the credibility of the pathway models.*

Four of our identified genes – *PIK3CA*, *KRAS*, *GNAS* and *ASXL1* – appear in Vogelstein's list of 'driver genes affected by subtle mutations' (the 20/20 list) [62]. Three genes – *PHOX2B*, *EXT1* and *RECQL4* – appear in Vogelstein's list of 'cancer predisposition genes', and one gene – *MYC* – appears in the list of 'genes affected by amplifications or deletions'. Thus, in total eight of 718 genes in our list appear in Vogelstein's lists, but our list is not *enriched* with Vogelstein's listed genes.

Our candidate genes are lethal upon inhibition in the context of DDR defects, as depicted by their relatively higher essentialities specifically in DDR-deficient cell lines. While some of our candidate genes constitute (universal) oncogenes (*e.g. KRAS* and *PIK3CA*), the roles of these genes can be understood only by considering the *context of DDR defects*. Depending on the status of certain DDR genes, our predicted genes can be oncogenic drivers (*e.g. TLK2*, which seems to be driving luminal breast tumours where *PIK3CA* expression is not high: **Section 3.4**), or vulnerability genes on which the survival of cells heavily depends (*e.g. SMC4*, which is required for structural maintenance of chromosomes and DNA repair, more so in the context of loss of key DDR genes).

A recent report has noted a context-dependent oncogenic role for *BRF2* (not a known oncogene in most cancers) in driving a subset of HER2+ breast tumours which do not express *ERRB2*/HER2 [63]; the authors call *BRF2* an 'alternative driver' of HER2+ tumours.

Therefore, although not enriched with genes from Vogelstein's 20/20 list of *universal* oncogenes, our predicted genes include oncogenes that are activated in a context-dependent manner in tumours. A general method based on the frequency of overexpressed/amplified genes will not be able to identify these genes; but by adding a context, we can prioritise them statistically. Our method addresses this by identifying oncogenes in the context of DDR defects.

*5. There exist many methods of inferring synthetic lethality (as reviewed by the authors) and the mutual exclusivity. Although the authors describe limits of previous methods inferring synthetic lethality, it would be relieving to see a reasonable overlap between the author's result and previous methods.*

The following oncogenes – *PIK3CA*, *MYC*, *EGFR* and *CCNE1* – from our list are known oncogenes and have also been predicted by earlier methods. These genes are also being pursued as targets in clinical trials. However, other predicted genes from the literature including *SMARCB1*, *MRE11A* and *ASPSCR1* [19] that have been inferred from lower-order organisms do not show up in our list. This is because these genes are infrequently altered in human cancers. As mentioned in **Discussion**, most of the genes predicted from morphologically simpler "model" organisms are highly conserved, and these are rarely altered in human cancers. Several of these genes are also tumour suppressors. As such, they are not worthwhile as anti-cancer targets.

**Additional file 1: Additional file 1.xlsx**

**File format:** .xlsx

**Title:** Additional file for 'Inferring synthetic lethal interactions from mutual exclusivity of genetic events in cancer'

**Description:** Additional results on predicted synthetic lethal genes

# Figures

**Figure 1**: Proportions of synthetic lethal (mutually exclusive) interactions identified for each of the six DDR genes *A* **(a)** across all cancers; and **(b)** in the individual cancers of breast, prostate and ovarian (uterine cancer by itself has too few samples to identify any significant interactions) at three levels of significance ($p < 0.01$, $0.001$ and $0.001$). Here, the SL partners *B* are amplified/upregulated.

**Figure 2**: Proportions of synthetic lethal (mutually exclusive) interactions identified for each of the six DDR genes *A* **(a)** across all cancers; and **(b)** in the individual cancers of breast, prostate and ovarian (uterine cancer by itself has too few samples to identify any significant interactions) at three levels of significance ($p < 0.05$ and $0.01$). Here, the SL partners *B* are deleted/downregulated.

**Figure 3:** Validation of synthetic lethal interactions against GARP essentiality scores from cell line screens [31,32]. The *left-hand plots* compare the ranges for GARP scores of our predicted genes *B* (amplified/upregulated) with that of the entire set (~16000) of profiled genes. While it is difficult to directly compare the two ranges because of the difference in the number of genes in them, for majority of the cell lines the gene *B* at the $25^{th}$ percentile had lower GARP scores than the corresponding gene from the entire profiled set. By $\chi^2$ test, genes *B* were significantly enriched ($p<10^{-5}$) within the top-quartile essential genes in these cell lines. The *right-hand plots* show GE ranks *vs* ME ranks for genes *B* in cell lines that are deficient in genes *A*: **(a)** *ATM*, **(b)** *BRCA1*, **(c)** *BRCA2*, **(d)** *PTEN* and **(e)** *TP53*.

**Figure 4:** Validation of synthetic lethal interactions against GARP essentiality scores from cell line screens [31,32]. The *left-hand plots* compare the ranges for GARP scores of our predicted genes *B* (deleted/downregulated) with that of the entire set (~16000) of profiled genes. While it is difficult to directly compare the two ranges because of the difference in the number of genes in them, for majority of the cell lines the gene *B* at the $25^{th}$ percentile had lower GARP scores than the corresponding gene from the entire profiled set. By $\chi^2$ test, genes *B* were significantly enriched ($p<10^{-5}$) with the top-quartile essential genes in these cell lines. The *right-hand plots* show GE ranks *vs* ME ranks for genes *B* in cell lines that are deficient in genes *A*: **(a)** *BRCA1*, **(b)** *BRCA2*, and **(c)** *PTEN*.

<span style="color:red">**Figure 5:** Differential essentiality of genes *B* **(a)** between nine DDR-deficient and MCF7 cell lines; and **(b)** between *PTEN*-/- and *PTEN* wild-type isogenic cell lines. We considered MCF7, which does not have any known DDR defect, as our control. Comparisons of GARP-score means for genes *B* between DDR-deficient lines and MCF7 showed significant differences (ANOVA $p<0.0001$) between these cell lines. Similarly, comparison of GARP</span>

scores between two isogenic HCT116-derived cell lines, one with *PTEN-/-* and the other with wild type *PTEN* showed significant difference (paired t-test: *p*<0.0001) between the two cell lines. This analysis indicated that the essentiality/lethality of genes *B* is specific to DDR-deficient/*PTEN*-deficient cell lines, and therefore context-dependent on DDR deficiency.

**Figure 6:** Snapshot of the proportion of cases and Kaplan-Meier survival plots for predicted genes *B* (amplified/upregulated) using survival data (untreated) from 1000 breast cancer patients, plotted using KMPlotter-breast (http://www.kmplot.com/ [48]). The patients are divided into two groups based on the overexpression (upper tertile of expression levels) and underexpression (below the upper tertile of expression levels) for these genes.

**Figure 7:** Two models for pathway-based targeting of synthetic lethal genes *B* in conjunction with deleted/downregulated genes *A*: **(a)** parallel pathways model where targeting *B* results in disruption of both survival pathways, and **(b)** negative feedback-loop model where targeting *B* shunts of (forward) signals for cell survival.

## Tables

| Cancer | Genomic | Gene expression | Total |
|---|---|---|---|
| Breast | 847 | 1182 | 2029 |
| Prostate | 152 | 471 | 623 |
| Ovarian | 562 | 266 | 828 |
| Uterine | 443 | 57 | 500 |
| *Total* | *2004* | *1976* | ***3980*** |

**Table 1: Number of genomic copy-number and gene-expression cancer samples used in the study.** The datasets were downloaded for four cancers – breast, prostate, ovarian and uterine – from CBioportal (http://www.cbioportal.org/index.do) [41,42] and TCGA Firehose (http://gdac.broadinstitute.org/), giving a total of 3980 cancer samples.

| Cell line / Gene *A* | *ATM* | *BRCA1* | *BRCA2* | *PTEN* | *TP53* |
|---|---|---|---|---|---|
| **HCC1500** | MUT; DOWN | | | | |
| **HCC1419** | DOWN | | | | MUT |
| **HCC1395** | DOWN | MUT; DOWN | MUT; | DOWN | MUT |
| **HCC1806** | | | | | |
| **HCC1143** | | | | | MUT |
| **HCC38** | | DOWN | | DOWN | MUT; HOMDEL |
| **HCC1187** | DOWN | DOWN | | | |
| **HCC1954** | DOWN | | | | MUT |
| **MCF7** | DOWN | | | | |
| **HCC1428** | DOWN | | | | HOMDEL |

**Table 2: Cell lines and the defects they harbour for DNA-damage response (DDR) genes.** Genome-wide (~16000 genes) essentiality data [31,32] from ten cancer cell lines that harbour defects (mutations MUT, downregulation DOWN or homologous deletions HOMDEL) in at least one of the DDR genes *ATM*, *BRCA1*, *BRCA2*, *PTEN* and *TP53*.

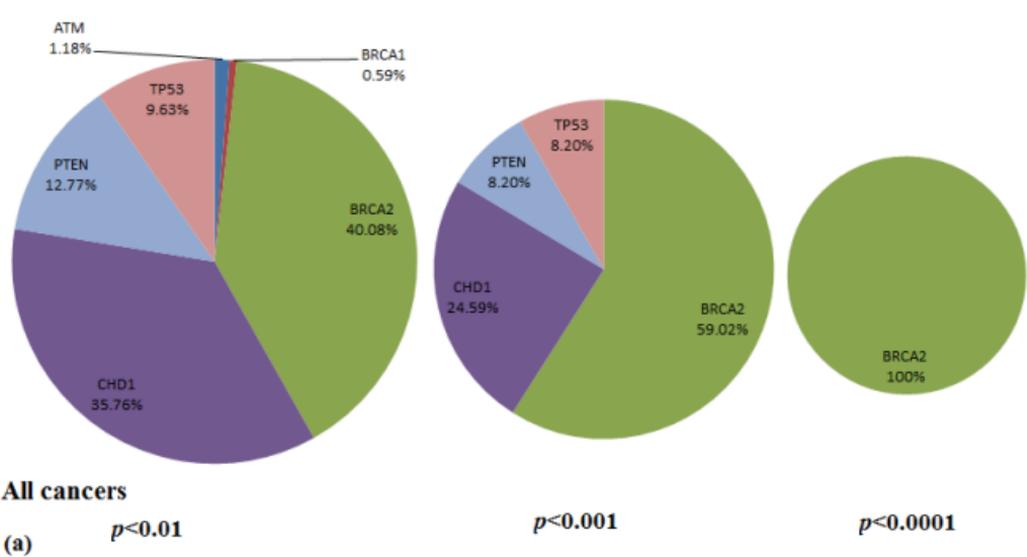
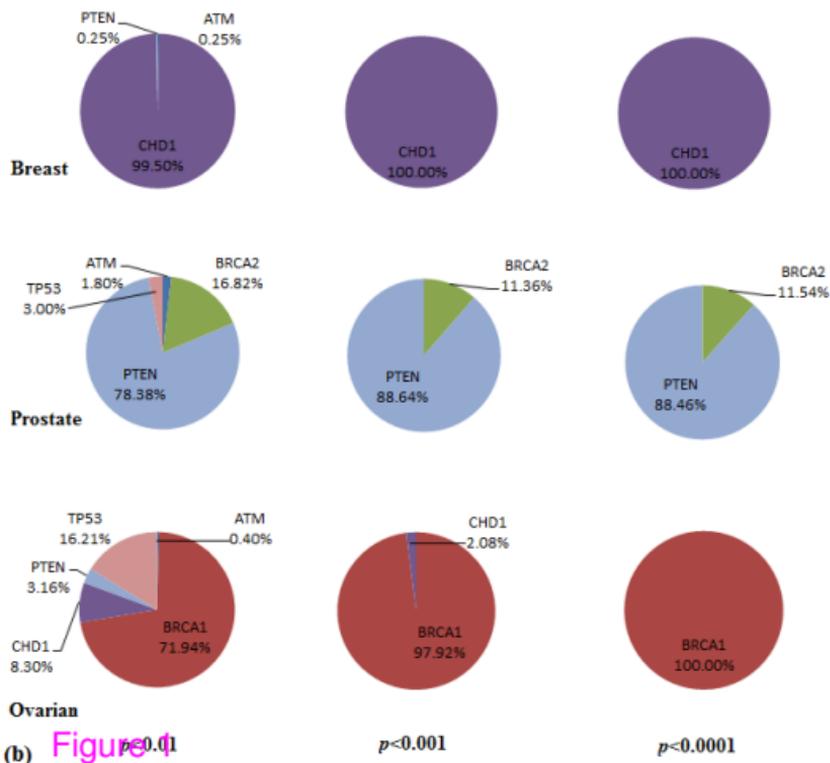

Figure 9.1

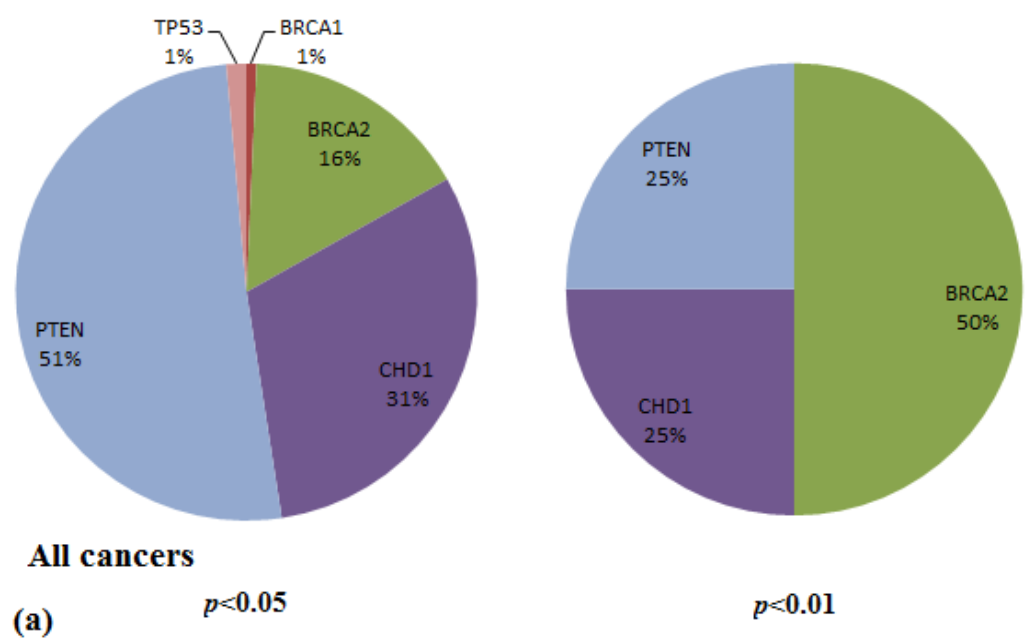

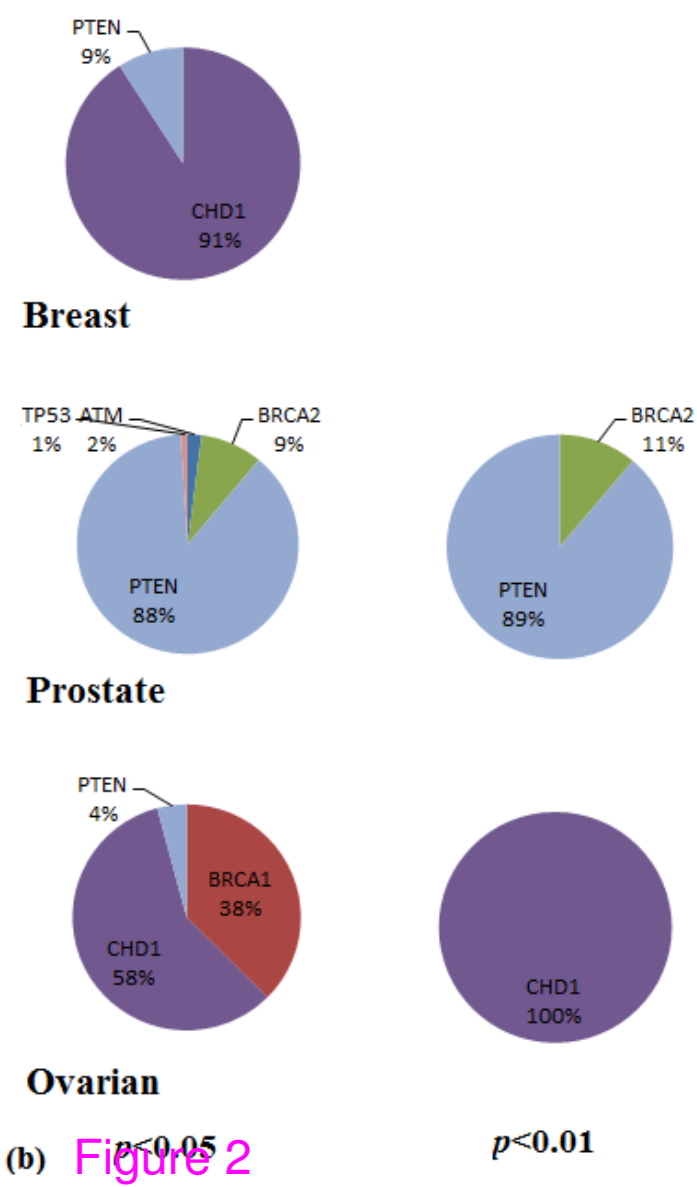

Figure 2

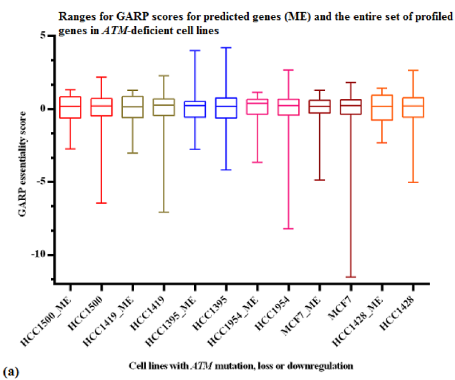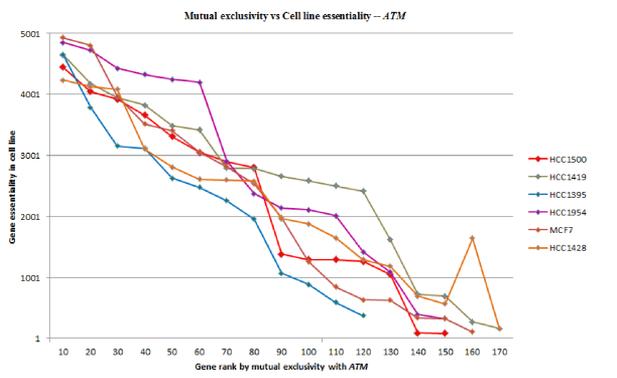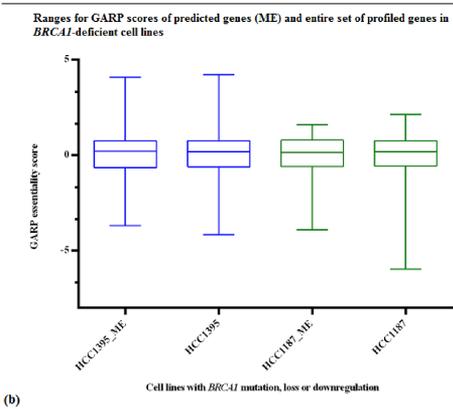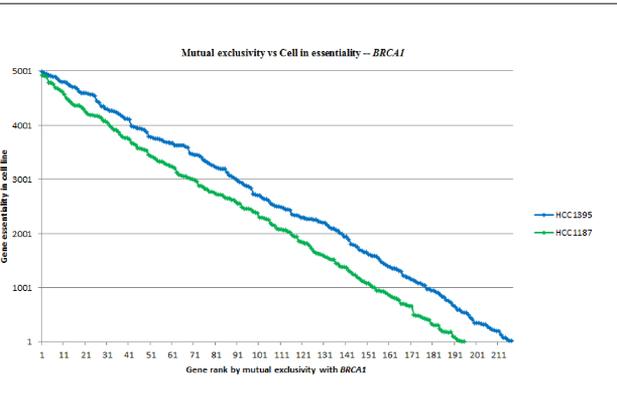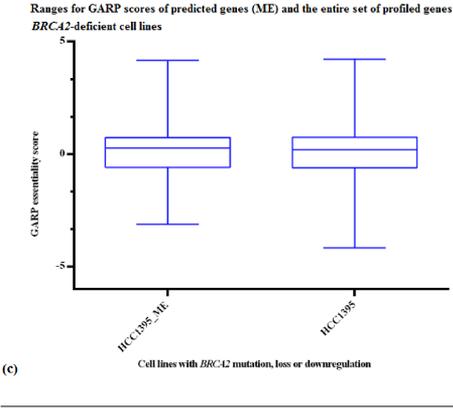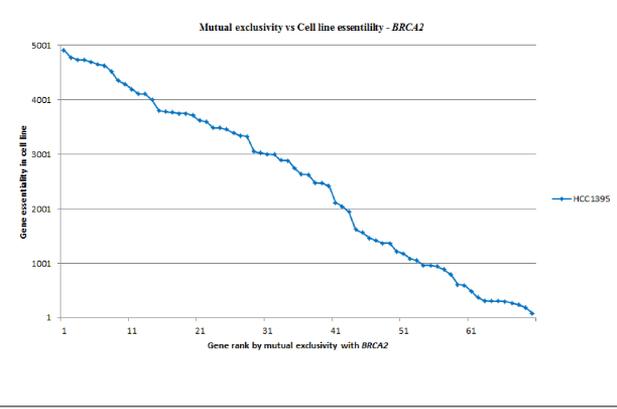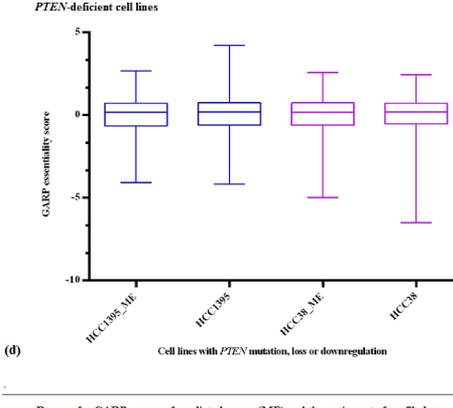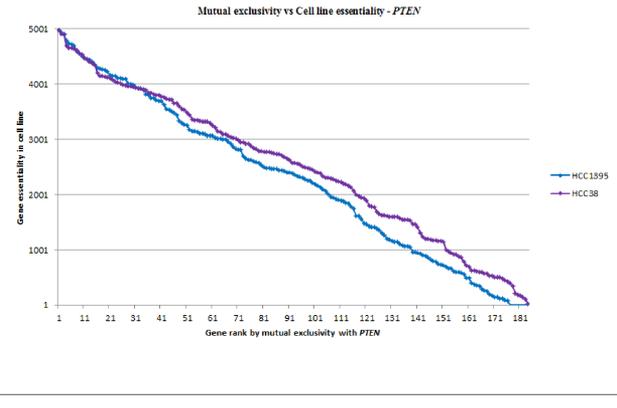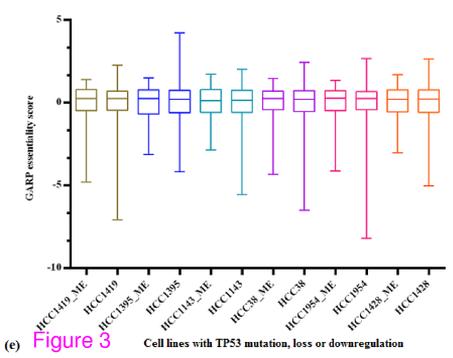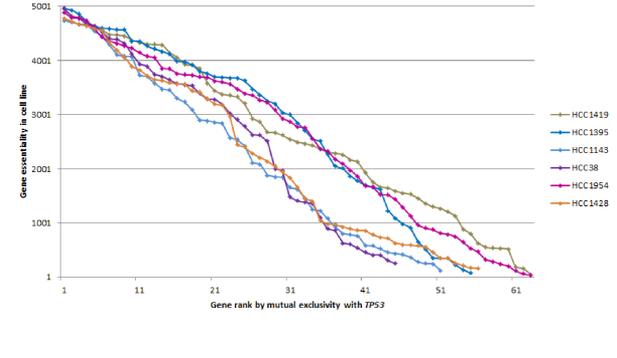

Figure 3

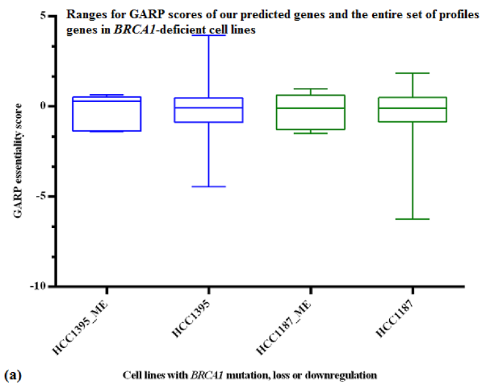
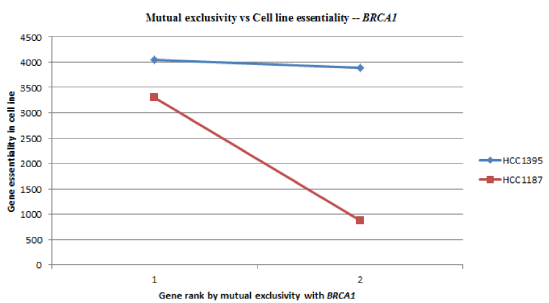

**(a)**

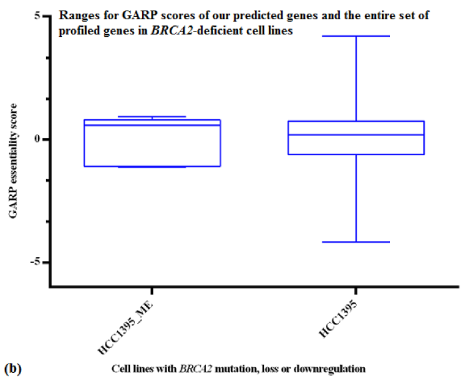
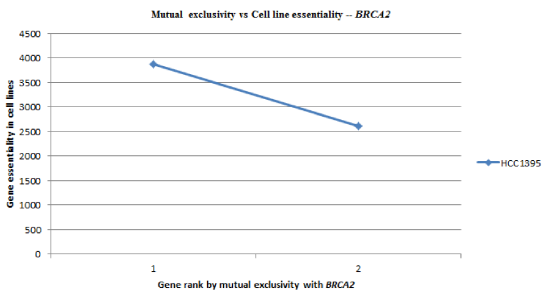

**(b)**

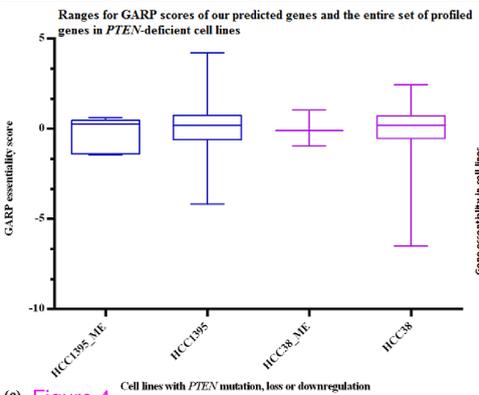
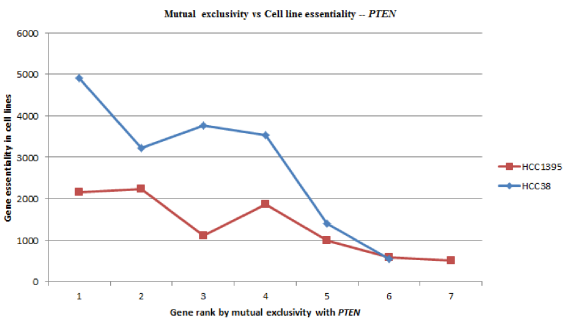

**(c)** Figure 4

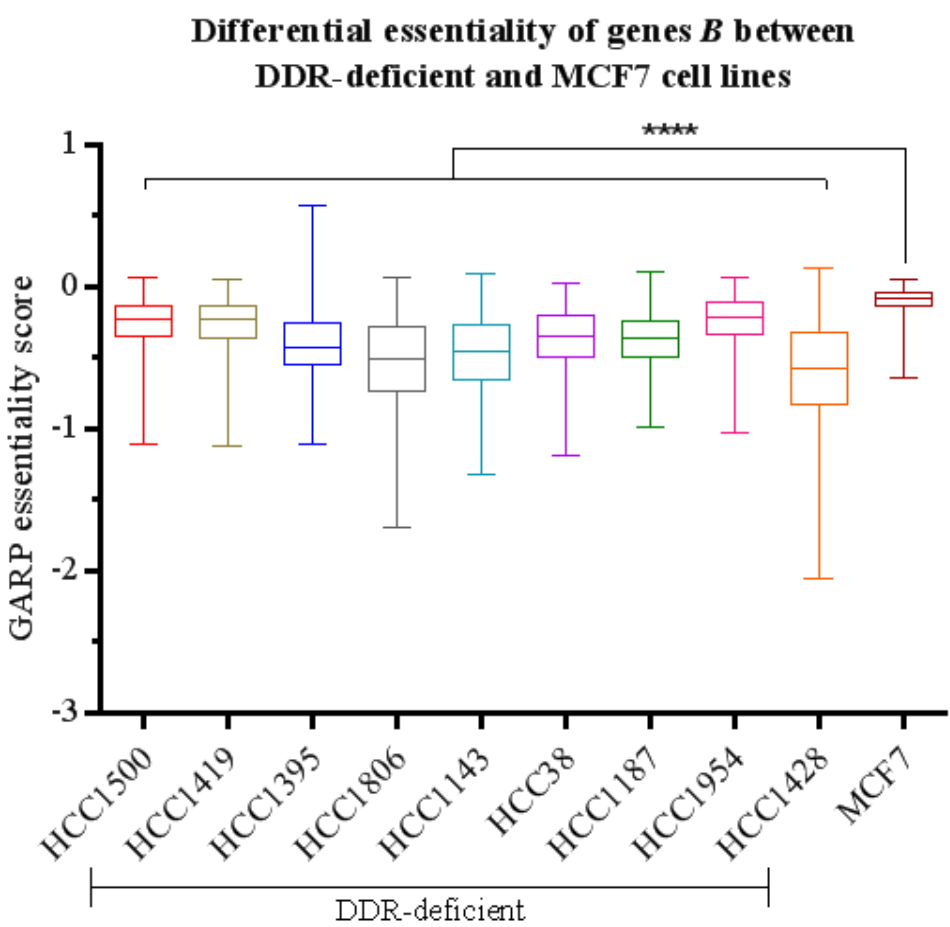 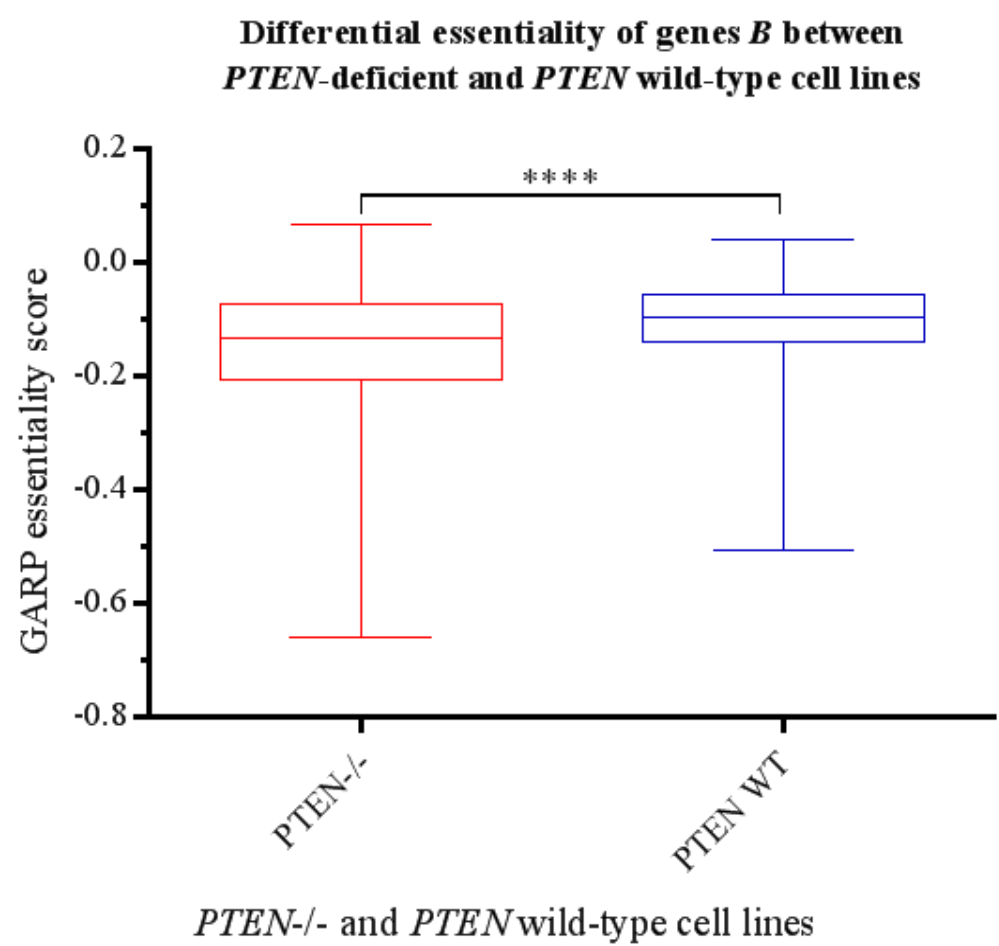

(a) Figure 5 DDR-deficient and MCF7 cell lines  (b)

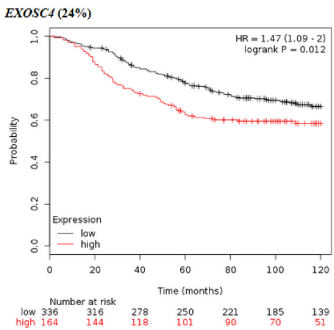
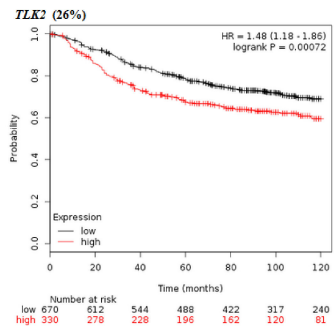
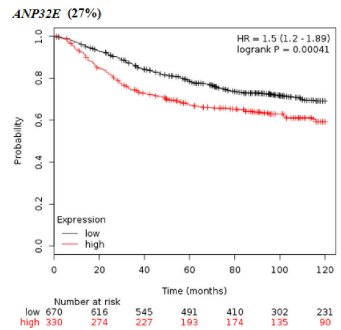
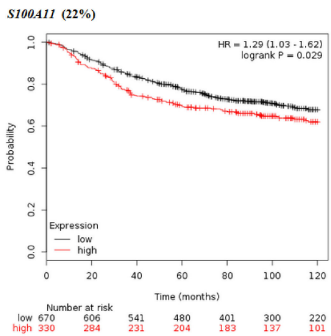
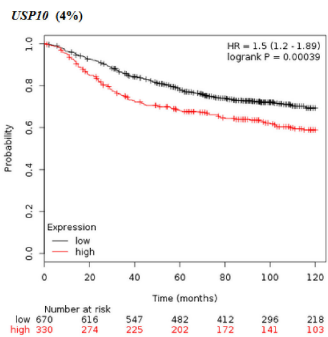
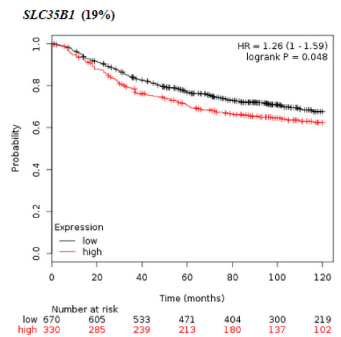
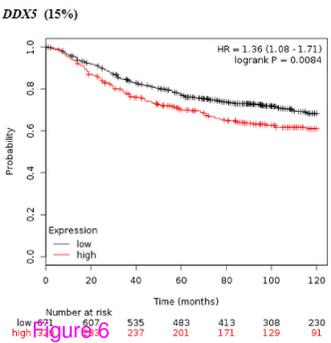
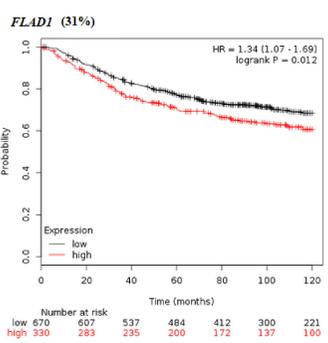
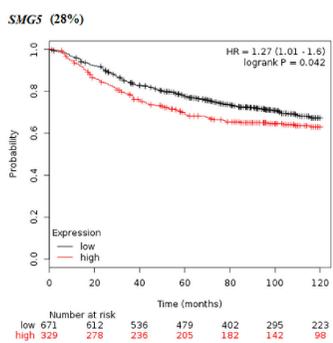

Figure 6

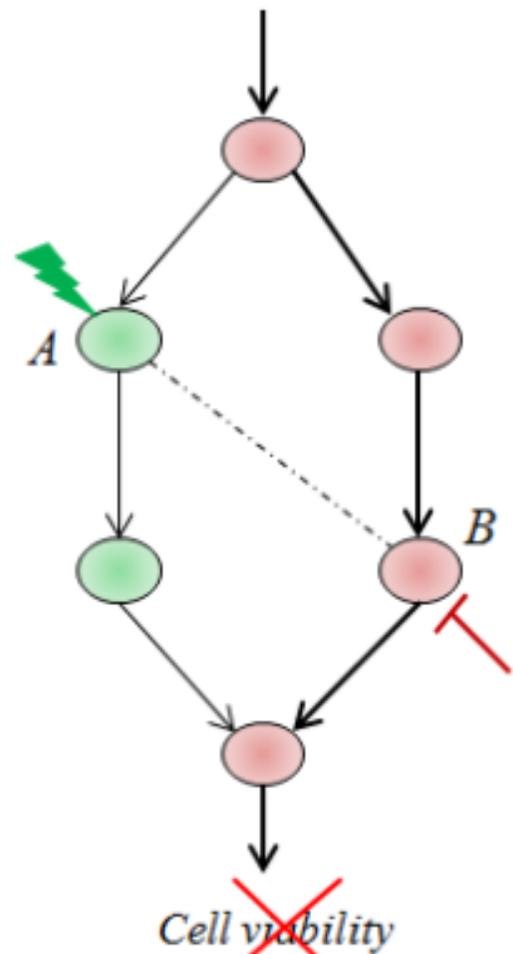 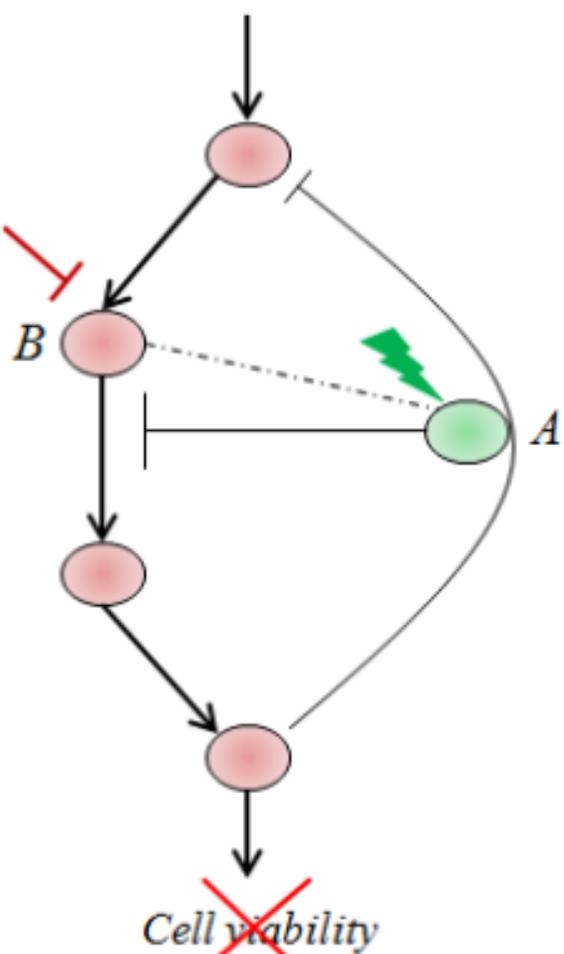

Figure 7
(a)  (b)

Cell viability

**Additional files provided with this submission:**

Additional file 1: Additional file 1.xlsx, 963K
http://www.biologydirect.com/imedia/1560412533187222/supp1.xlsx